\documentclass[twocolumn,amsmath,amssymb,showpacs]{revtex4}
\usepackage{bm}
\usepackage{graphicx}
\usepackage{epstopdf}

\begin{document}

\newcommand{\uu}[1]{\underline{#1}}
\newcommand{\pp}[1]{\phantom{#1}}
\newcommand{\be}{\begin{eqnarray}}
\newcommand{\ee}{\end{eqnarray}}
\newcommand{\ve}{\varepsilon}
\newcommand{\vs}{\varsigma}
\newcommand{\Tr}{{\,\rm Tr\,}}
\newcommand{\pol}{\frac{1}{2}}
\newcommand{\CC}{\rotatebox[origin=c]{180}{$\mathbb{C}$} }
\newcommand{\Exp}{{\,\rm Exp\,}}
\newcommand{\Sin}{{\,\rm Sin\,}}
\newcommand{\Cos}{{\,\rm Cos\,}}

\title{
Fourier transforms on Cantor sets: A study in non-Diophantine arithmetic and calculus}
\author{Diederik Aerts$^1$, Marek Czachor$^{1,2}$, and Maciej Kuna$^{1,2}$}
\affiliation{
$^1$ Centrum Leo Apostel (CLEA),
Vrije Universiteit Brussel, 1050 Brussels, Belgium,\\
$^2$ Wydzia{\l} Fizyki Technicznej i Matematyki Stosowanej,
Politechnika Gda\'nska, 80-233 Gda\'nsk, Poland
}

\begin{abstract}
Fractals equipped with intrinsic arithmetic lead to a natural definition of differentiation, integration and complex numbers. Applying the formalism to the problem of a Fourier transform on fractals we show that the resulting transform has all the expected basic properties. As an example we discuss a sawtooth signal on the ternary middle-third Cantor set. The formalism works also for fractals that are not self-similar.
\end{abstract}
\pacs{04.60.-m, 05.45.Df, 02.20.Qs}
\maketitle

\section{Introduction}

While trying to formulate quantum mechanics on fractal backgrounds one immediately faces the problem of momentum representation. The issue is nontrivial and reduces to the question of what should be meant by a Fourier transform on a fractal. Historically the first approach to fractal harmonic analysis can be, implicitly, traced back to studies of diffusion on fractals \cite{Kusuoka,Goldstein}. A generator of the diffusion is then a candidate for a Laplacion on a fractal, and once we have a Laplacian we can look for its eigenfunctions. The eigenfunctions may play a role of a Schauder basis in certain function spaces, and thus lead to a sort of signal analysis on a fractal. Whether and under what conditions the resulting eigenfunction expansions can be regarded as analogs of Fourier transformations is a separate story. Fractals such as Cantor sets naturally lead to wavelet transforms (the Haar basis \cite{Semadeni,Ciesielski,J}, for example), but quantum mechanical momentum representation is expected to be associated with gradient operators, and there is no obvious link between Haar wavelets and gradients.

Gradients and Laplacians can be defined on fractals also more directly. Here one should mention the approaches that begin with Dirichlet forms defined on certain self-similar fractals, and those that start with discrete Laplacians \cite{Kusuoka1989,Kigami1989,Kigami,Strichartz}. Self-similarity is an important technical assumption, and it is not clear what to do in more realistic cases, such as multi-fractals or fractals that have no self-similarity at all (a generic case in natural systems). Four different definitions of a gradient (due to Kusuoka, Kigami, Strichartz and Teplyaev) can be found in \cite{Tep}.

One might naively expect that it would be more logical to begin with first derivatives and only then turn to higher-order operators, such as Laplacians. It turns out that Laplacians defined in the above ways cannot be regarded as second-order operators. Still, an approach where Laplacians are indeed second-order is possible and was introduced by Fujita \cite{Fujita1987,Fujita}, and further developed by Freiberg, Z\"ahle and others \cite{FZ,Z,F,Arzt,Kess}. We will later see that a non-Diophantine Laplacian is exactly second-order and, similarly to the approach from \cite{FZ,Z,F,Arzt,Kess}, is based on derivatives and integrals satisfying the fundamental laws of calculus.

In yet another traditional approach to harmonic analysis on fractals one begins with self-similar fractal measures, and then seeks exponential functions that are orthogonal and complete with respect to them. The classic result of Jorgensen and Pedersen \cite{JP} states that such exponential functions do exist on certain fractals, such as the quaternary Cantor set, but are excluded in the important case of the ternary middle-third Cantor set.

In the present paper we will follow a different approach. One begins with arithmetic operations (addition, subtraction, multiplication, and division) which are intrinsic to the fractal. The arithmetic so defined is non-Diophantine in the sense of Burgin \cite{Burgin1,Burgin2}. An important step is then to switch from arithmetic to calculus \cite{MC} where, in particular, derivatives and integrals are naturally defined. The resulting formalism is simple and general, extends beyond fractal applications, but works with no difficulty for Cantorian fractals, even if they are not self-similar \cite{MC,ACK}. Actually, a straightforward motivation for the present paper came from discussions with the referee of \cite{ACK}, who pointed out possible difficulties with momentum representation of quantum mechanics on Cantorian space-times.

In Sec.II we recall the basic properties of non-Diophantine arithmetic, illustrated by four examples from physics, cognitive science, and fractal theory.
Sec. III is devoted to complex numbers, discussed along the lines proposed by one of us in \cite{MC}, and with particular emphasis on trigonometric and exponential functions. In Sec. IV we recall the non-Diophantine-arithmetic definitions of derivatives and integrals. Sec. V discusses a scalar product of functions, and the corresponding Fourier transform (both complex and real) is introduced in Sec. VII. In Sec. VIII we discuss an explicit example of a sawtooth signal with Cantorian domain and range. Finally, in Sec. IX we briefly discuss the issue of spectrum of Fourier frequencies, and compare our results with those from \cite{JP}.

\section{Generalized arithmetic: Fractal and not only}

Consider a set $\mathbb{X}$ and a bijection $f:\mathbb{X} \to \mathbb{R}$
Following the general formalism from \cite{MC} we define the arithmetic operations in $\mathbb{X}$,
\be
x\oplus y &=& f^{-1}\big(f(x)+f(y)\big),\nonumber\\
x\ominus y &=& f^{-1}\big(f(x)-f(y)\big),\nonumber\\
x\odot y &=& f^{-1}\big(f(x)f(y)\big),\nonumber\\
x\oslash y &=& f^{-1}\big(f(x)/f(y)\big),\nonumber
\ee
for any $x,y\in \mathbb{X}$. In later applications we will basically concentrate on an appropriately constructed fractal $\mathbb{X}$, but the results are more general. This is an example of a non-Diophantine arithmetic \cite{Burgin1,Burgin2}.

One verifies the standard properties: (1) associativity $(x\oplus y)\oplus z=x\oplus (y\oplus z)$,
$(x\odot y)\odot z=x\odot (y\odot z)$, (2) commutativity $x\oplus y=y\oplus x$, $x\odot y=y\odot x$, (3) distributivity
$(x\oplus y)\odot z=(x\odot z)\oplus (y\odot z)$. Elements $0',1'\in X$ are defined by $0'\oplus x=x$, $1'\odot x=x$, which implies $f(0')=0$, $f(1')=1$.
One further finds $x\ominus x=0'$, $x\oslash x=1'$, as expected.
A negative of $x\in \mathbb{X}$ is defined as $\ominus x=0'\ominus x=f^{-1}\big(-f(x)\big)$, i.e.
$f(\ominus x)=-f(x)$ and $f(\ominus 1')=-f(1')=-1$, i.e. $\ominus 1'=f^{-1}(-1)$.
Notice that
\be
(\ominus 1')\odot(\ominus 1')
&=&
f^{-1}\left(f(\ominus 1')^2\right)=f^{-1}(1)=1'.
\ee
Multiplication can be regarded as repeated addition in the following sense. Let $n\in\mathbb{N}$ and $n'=f^{-1}(n)\in \mathbb{X}$. Then
\be
n'\oplus m'
&=&
(n+m)',\\
n'\odot m'
&=&
(nm)'\\
&=&
\underbrace{m'\oplus\dots \oplus m'}_{n\rm{ times}}.
\ee
In particular
$n'=1'\oplus\dots \oplus 1'$ ($n$ times).

A power function $A(x)=x\odot\dots \odot x$ ($n$ times) will be denoted by $x^{n'}$. Such a notation is consistent in the sense that
\be
x^{n'}\odot x^{m'}=x^{(n+m)'}=x^{n'\oplus m'}.
\ee
Before we plunge into fractal applications let us consider four explicit examples of non-Diophantine arithmetic.

\subsection{Benioff's number scaling}

The rescaled-multiplication approach of Benioff \cite{Benioff,Benioff2} can be regarded as a particular case of the above formalism with $f(x)=px$, $p\neq 0$. Indeed, $x\odot y=(1/p)(pxpy)=pxy$, $x\oplus y=(1/p)(px+py)=x+y$, $x\oslash y=(1/p)(px)/(py)=x/(py)$, but $f(1/p)=1$. Since
$(1/p)\odot x=(1/p)\big(p(1/p)px\big)=x$ one infers that $1'=f^{-1}(1)=1/p$ is the unit element of multiplication in Benioff's non-Diophantine arithmetic.

\subsection{Fechner map}

This arithmetic is implicitly used in cognitive science \cite{MC1}. It occurs as a solution of the following Weber--Fechner problem \cite{BairdNoma}: Find a generalized arithmetic such that
$(x+kx)\ominus x$ is independent of $x$. Here $x\mapsto x'=x+\Delta x$ is the change of an input signal, while $x'\ominus x$ is the change of $x$ as perceived by a nervous system. Experiments show that $\Delta x/x\approx k= \textrm{const}$ (Weber-Fechner law) in a wide range of $x$s, and with different values of $k$ for different types of stimuli. The corresponding arithmetic is defined by the `Fechner map' $f(x)=a\ln x+b$, $f^{-1}(x)=e^{(x-b)/a}$, and thus $0'=f^{-1}(0)=e^{-b/a}$, $1'=f^{-1}(1)=e^{(1-b)/a}$. Clearly, $0'\neq 0$ and $1'\neq 1$. Interestingly, the Fechnerian negative of $x\in \mathbb{R}_+$ reads
\be
\ominus x
&=&
0'\ominus x
=e^{-2b/a}/x \in \mathbb{R}_+,
\ee
but nevertheless does satisfy
\be
\ominus x\oplus x
&=&
e^{-b/a}
=0',
\ee
as it should on general grounds \cite{MC1}. So, numbers that are negative with respect to one arithmetic are positive with respect to another. In a future work we will show that Fechner's $f$ has intriguing consequences for relativistic physics.

\subsection{Ternary Cantor line}

Let us start with the right-open interval $[0,1)\subset \mathbb{R}$, and let the (countable) set $\mathbb{Y}_2\subset[0,1)$ consist of those numbers that have two different binary representations. Denote by $0.t_1t_2\dots$ a ternary representation of some $x\in [0,1)$. If $y\in \mathbb{Y}_1=[0,1)\setminus \mathbb{Y}_2$ then $y$ has a unique binary representation, say $y=0.b_1b_2\dots$. One then sets $g_\pm(y)=0.t_1t_2\dots$, $t_j=2b_j$. The index $\pm$ appears for the following reason. Let $y=0.b_1b_2\dots=0.b'_1b'_2\dots$ be the two representations of $y\in \mathbb{Y}_2$. There are two options, so we define:  $g_-(y)=\min\{0.t_1t_2\dots,0.t'_1t'_2\dots\}$ and $g_+(y)=\max\{0.t_1t_2\dots,0.t'_1t'_2\dots\}$, where $t_j=2b_j$, $t'_j=2b'_j$. We have therefore constructed two injective maps $g_\pm:[0,1)\to [0,1)$. The ternary Cantor-like sets are defined as the images $C_{\pm}(0,1)=g_\pm\big([0,1)\big)$, and $f_\pm:C_{\pm}(0,1)\to[0,1)$, $f_\pm=g_\pm^{-1}$, is a bijection between $C_{\pm}(0,1)$ and the interval. For example, $1/2\in \mathbb{Y}_2$ since $1/2=0.1_2=0.0(1)_2$. We find
\be
g_-(1/2) &=&\min\{0.2_3=2/3,0.0(2)_3=1/3\}=1/3,\\
g_+(1/2) &=&\max\{0.2_3=2/3,0.0(2)_3=1/3\}=2/3.
\ee
Accordingly, $1/3\in C_-(0,1)$ while $2/3\notin C_-(0,1)$. And vice versa, $1/3\notin C_+(0,1)$, $2/3\in C_+(0,1)$. The standard Cantor set is the sum $\tilde C=C_-(0,1)\cup C_+(0,1)$. All irrational elements of $\tilde C$ belong to $C_\pm(0,1)$ (an irrational number has a unique binary form), so $\tilde C$ and $C_\pm(0,1)$ differ on a countable set. Notice further that $0\in C_\pm(0,1)$, with $f_\pm(0)=0$. In \cite{MC,ACK} we worked with $C_-(0,1)$ so let us concentrate on this case. Let $C_-(k,k+1)$, $k\in \mathbb{Z}$, be the copy of $C_-(0,1)$ but shifted by $k$. We construct a fractal $\mathbb{X}=\cup_{k\in\mathbb{Z}}C_-(k,k+1)$, and the bijection $f:\mathbb{X}\to \mathbb{R}$. Explicitly, if $x\in C_-(0,1)$, then $x+k\in C_-(k,k+1)$, and $f(x+k)=f(x)+k$ by definition. In \cite{MC,ACK} the set $\mathbb{X}$ is termed the Cantor line, and $f$ is the Cantor-line function. For more details see \cite{MC}. The set $\mathbb{X}\cap [k,k+1)$ is self-similar, but $\mathbb{X}$ as a whole is not-self similar. Fig.~1 (upper) shows the plot of $g=f^{-1}$. For completely irregular generalizations of the Cantor line, see \cite{ACK}.

Let us make a remark that in the literature one typically considers Cantor sets $\tilde C$ so that the resulting function $g:\tilde C\to [0,1)$ is non invertible on a countable subset. In \cite{Semadeni} one employs the map $g$ to define the Haar basis on $\tilde C$ `up to a countable set of points'. In our formalism we have to work with bijective $g$ since we need its inverse.

\subsection{Quaternary Cantor line}

Here we construct a Cantor set that is analogous to the one employed by Jorgensen and Pedersen in \cite{JP}. In a single step of the algorithm one splits an interval into four identical segments and retains only the first and the third. Similarly to the triadic set one needs to remove a countable subset of right or left endpoints of the sub-intervals in order to have a one-to-one map onto $[0,1)$. We then extend the construction in a self-similar way to the whole of $\mathbb{R}$. So, as opposed to the previous paragraph, we will not consider the sum of translated copies, but rather the sum of rescaled copies.

Consider a number $y\in \mathbb{R}_+$. Let $\mathbb{Y}_1$ denote those $y$ that have a unique binary representation
\be
y=(b_m\dots b_1b_0.b_{-1}\dots b_{-k}\dots)_2.
\ee
We define
\be
g_\pm(y)=(2b_m\dots 2b_12b_0.2b_{-1}\dots 2b_{-k}\dots)_4.\label{quat}
\ee
$g_\pm(y)$ is a number whose quaternary (i.e. base-four) representation contains only 0s and 2s. The ternary set had the same property, but in the ternary (base-three) representation.

Now, if $y\in\mathbb{Y}_2=\mathbb{R}\setminus \mathbb{Y}_1$ we have the ambiguity which of the two binary forms of $y$ to take.
Applying to the two forms the recipe (\ref{quat}) we get two numbers, $x$ and $x'$ say. Then
$g_-(y)=\min\{x,x'\}$ and $g_+(y)=\max\{x,x'\}$. Finally, we extend the maps by anti-symmetry to negative $y$, i.e. $g_\pm(-y)=-g_\pm(y)$.

The images $\mathbb{X}_\pm=g_\pm(\mathbb{R})$ define two quaternary Cantor sets, and $f_\pm=g_\pm^{-1}$ are the required bijections $f_\pm:\mathbb{X}_\pm\to\mathbb{R}$.

Fig.~1 (lower) shows the plot of $g_+$. Notice that
\be
1_+' &=&
f_+^{-1}(1)=g_+\big(1.(0)_2\big)=g_+\big(0.(1)_2\big)
\nonumber\\
&=&
\max\{2.(0)_4=2,0.(2)_4=2/3\}=2.
\ee
Analogously
\begin{figure}\label{Fig1}
\includegraphics[width=8 cm]{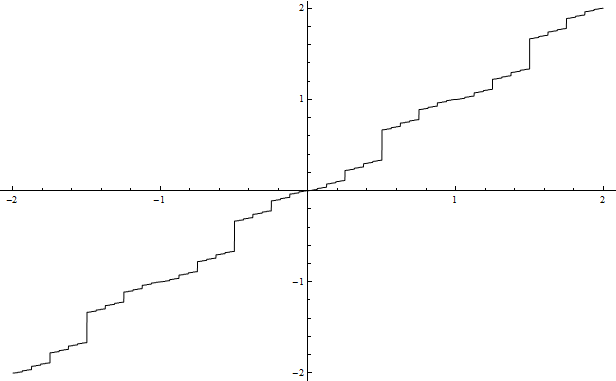}
\includegraphics[width=8 cm]{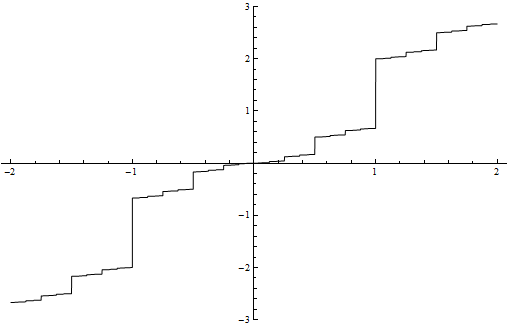}
\caption{$f^{-1}$ for the ternary Cantor line $\mathbb{X}$ (upper), and $f^{-1}_+$ for the quaternary Cantor set $\mathbb{X}_+=f^{-1}_+(\mathbb{R})$ (lower).}
\end{figure}
\be
1_-' &=&
f_-^{-1}(1)=g_-\big(1.(0)_2\big)=g_-\big(0.(1)_2\big)
\nonumber\\
&=&
\min\{2.(0)_4=2,0.(2)_4=2/3\}=2/3.
\ee
As we can see, both unit elements $1_\pm'$ differ from 1, so it is not clear which of the two bijections, and thus which of the two Cantor sets $\mathbb{X}_\pm$, is more `natural'. Calculations are simpler with $\mathbb{X}_+$.

\section{Non-Diophantine complex numbers}

The examples discussed in the present paper will employ real-valued, sine and cosine Fourier transforms. However, having in mind future applications it will pay to discuss in detail the construction of a complex-valued transform. In order to do so, we have to explain what should be meant by a complex number if non-Diophantine arithmetic is in use. We will follow the strategy from \cite{MC}.

From now on the numbers from $\mathbb{X}$ will be denoted by upper-case letters: $X\in\mathbb{X}$, $X\oplus Y\in\mathbb{X}$, and so on. The elements of $\mathbb{R}$ will be generally denoted by lower-case symbols, eg. $f(X)=x$, with very few non-ambiguous exceptions, such as $n!'=f^{-1}(n!)$ instead of the apparently more consistent $N!'=f^{-1}(n!)$. Non-Diophantine complex numbers, denoted by $\CC$, will be identified with pairs of elements form $\mathbb{X}$, subject to the following arithmetic:
\be
A\oplus B
&=&
(A_1,A_2)\oplus(B_1,B_2)\\
&=&(A_1\oplus B_1,A_2\oplus B_2),\label{c oplus}\\
A\odot  B
&=&
(A_1,A_2)\odot (B_1,B_2)\\
&=&
(A_1\odot  B_1\ominus  A_2\odot  B_2,A_1\odot  B_2\oplus A_2\odot  B_1),\label{c odot}
\ee
and conjugation
\be
A^* &=& (A_1,\ominus  A_2).\label{conj}
\ee
The modulus is defined by
\be
|A|^{2'}
&=&
A\odot  A^*=
(A_1^{2'}\oplus A_2^{2'},0')\equiv A_1^{2'}\oplus A_2^{2'}.
\ee
We simplify the notation by identifying $(A_1,0')\in \CC$ with $A_1\in \mathbb{X}$.

The `imaginary unit' is defined as $i'=(0',1')$, and satisfies $i'\odot i'=i'^{2'}=(\ominus 1',0')\equiv\ominus 1'$.
We do not risk any ambiguity if we write $i'A$ instead of $i'\odot A$, for any $A\in\CC$.

Moreover
\be
A_1\oplus i' A_2
&=&
(A_1,0')\oplus i'(A_2,0')\nonumber\\
&=&
(A_1,0')\oplus (0',1')\odot  (A_2,0')\nonumber\\
&=&
(A_1,0')\oplus (0',A_2)=(A_1,A_2).
\ee
Complex exponent is defined as
\be
\Exp(i'\phi)
&=&(\Cos\phi,\Sin\phi)\\
&=&
\Cos\phi \oplus i'\Sin\phi,
\\
&=&
f^{-1}\Big(\cos f(\phi)\Big)\oplus i' f^{-1}\Big(\sin f(\phi)\Big)\\
&=&
f^{-1}\Big(\Re e^{if(\phi)}\Big)\oplus i' f^{-1}\Big(\Im e^{if(\phi)}\Big)
\ee
where
\be
\Cos X &=& f^{-1}\big(\cos f(X)\big),\label{Cos}\\
\Sin X &=& f^{-1}\big(\sin f(X)\big)\label{Sin}.
\ee
The trigonometric identity reads
\be
1' &=&\Cos^{2'}X\oplus \Sin^{2'}X\\
&=&
\Exp(i'\phi)\odot \Exp(i'\phi)^* \\
&=&
\Exp(i'\phi)\odot \Exp(\ominus i'\phi).
\ee
In Taylor expansions we need a non-Diophantine factorial
\be
n!'
&=&
1'\odot 2'\odot 3'\dots \odot n'\\
&=&
f^{-1}\Big(f(1')f(2')f(3')\dots f(n')\Big)\\
&=&
f^{-1}(1\cdot 2\dots \cdot n)=f^{-1}(n!).
\ee
Taylor expansions of elementary functions occur automatically,
\be
\Cos X &=& f^{-1}\big(\cos f(X)\big)\\
&=&
f^{-1}\Big(1-f(X)^2/2!+f(X)^4/4!-\dots\Big)\\
&=&
f^{-1}\Big(f(1')-f(X)^2/f(2!')+\dots\Big)\\
&=&
1'\ominus X^{2'}\oslash 2!'\oplus X^{4'}\oslash 4!'\dots\\
&=&
\oplus_{k=0}^\infty (\ominus 1')^{(2k)'}X^{(2k)'}\oslash (2k)!'\\
\Sin X &=& f^{-1}\big(\sin f(X)\big)\\
&=&
X\ominus X^{3'}\oslash 3!'\oplus X^{5'}\oslash 5!'\dots\\
&=&
\oplus_{k=0}^\infty (\ominus 1')^{(2k)'}X^{(2k+1)'}\oslash (2k+1)!'\\
\Exp X
&=& f^{-1}\big(\exp f(X)\big)\\
&=&
1'\oplus X\oplus  X^{2'}\oslash 2!'\oplus X^{3'}\oslash 3!'\dots\\
&=&
\oplus_{k=0}^\infty X^{k'}\oslash k!'.
\ee

\section{Non-Diophantine derivatives and integrals}

A derivative of a function $A:\mathbb{X}\to \mathbb{X}$ is defined by
\be
\frac{DA(X)}{DX}
&=&
\lim_{H\to 0'}\big(A(X\oplus H)\ominus A(X)\big)\oslash H,\label{DA/DX}
\ee
and an integral is an inverse of the derivative, so that the fundamental laws of calculus relating integration and differentiation remain valid in $\mathbb{X}$.

For example, let $A(X)=X^{N'}=f^{-1}\big(f(X)^N\big)$. Directly from definition (\ref{DA/DX}), and taking into account $f(N')=N$, one finds
\be
\frac{DX^{N'}}{DX}
&=&
f^{-1}\big(Nf(X)^{N-1}\big)
\\
&=&
f^{-1}\big(f(N')f(X)^{N-1}\big)
\\
&=&
N'\odot X^{(N-1)'}=N'\odot X^{N'\ominus 1'}.
\ee
One similarly verifies
\be
\frac{D \Sin(K\odot X)}{DX}
&=&
K\odot \Cos (K\odot X),\\
\frac{D \Cos(K\odot X)}{DX}
&=&
\ominus K\odot \Sin (K\odot X),\\
\frac{D \Exp(K\odot X)}{DX}
&=&
K\odot \Exp (K\odot X),\\
\frac{D \Exp(i'K\odot X)}{DX}
&=&
i'K\odot \Exp (i'K\odot X).
\ee
A derivative of a function $a:\mathbb{R}\to \mathbb{R}$ is defined with respect to the lowercase arithmetic,
\be
\frac{da(x)}{dx}=\lim_{h\to 0}\big(a(x+h)-a(x)\big)/h.
\ee

Now let $A=f^{-1}\circ a\circ f$. Then,
\be
\frac{DA(X)}{DX}
&=&
f^{-1}\left(\frac{da\big(f(X)\big)}{df(X)}\right),\label{der}\\
\int_X^YA(X')DX'
&=&
f^{-1}\left(\int_{f(X)}^{f(Y)}a(x)dx\right),\label{int}
\ee
satisfy
\be
\frac{D}{DX}\int_{Y}^X A(X')DX' &=& A(X),\label{calc1}\\
\int_Y^X \frac{DA(X')}{DX'}DX' &=& A(X)\ominus A(Y).\label{calc2}
\ee
Formula (\ref{der}) follows directly from the definitions of $D/DX$ and $d/dx$.

It is is extremely important to realize that (\ref{der}) is {\it not\/} the usual formula relating derivatives of $A=f^{-1}\circ a\circ f$ and $a$. Indeed, \be
\frac{DA}{DX}  &=& f^{-1}\circ \frac{da}{dx} \circ f,\label{der'},
\ee
so that $D/DX$ behaves like a covariant derivative, but with a trivial connection. Yet, $f$ can be {\it any\/} bijection $f: \mathbb{X}\to \mathbb{R}$. The usual approach, employed in differential geometry or gauge theories, would employ the arithmetic of $\mathbb{R}$, and one would have to assume differentiability of $f$ and $f^{-1}$. Here bijectivity is enough since no derivatives of either $f$ or $f^{-1}$ will occur in (\ref{der}) and (\ref{der'}).

\section{Scalar product}

Let $A_k,B_k:\mathbb{X}\to \mathbb{X}$, $k=1,2$, $A_k=f^{-1}\circ a_k\circ f$, $B_k=f^{-1}\circ b_k\circ f$, and $A=A_1\oplus i'A_2$, $a=a_1+ia_2$.
Define
\be
\langle A|B\rangle
&=&
\int_{\ominus T\oslash 2'}^{T\oslash 2'} A(X)^*\odot B(X)DX.\label{<A|B>}
\ee
Employing (\ref{c odot}), (\ref{conj}), (\ref{int}) we transform (\ref{<A|B>}) into
\be
\langle A|B\rangle
&=&
f^{-1}\left(\Re\int_{-f(T)/2}^{f(T)/2}
\overline{a(x)}b(x)
dx\right)
\nonumber\\
&\pp=&
\oplus
i'f^{-1}\left(\Im\int_{-f(T)/2}^{f(T)/2}
\overline{a(x)}b(x)
dx\right).
\ee
$f(T)$ can be finite or infinite. It is useful to denote $\langle a|b\rangle=\int_{-f(T)/2}^{f(T)/2}\overline{a(x)}b(x)dx$, so that
\be
\langle A|B\rangle
&=&
f^{-1}\left(\Re\langle a|b\rangle\right)
\oplus
i'f^{-1}\left(\Im\langle a|b\rangle\right).
\ee

In the Appendix we prove that
\be
\langle A|B\rangle^*
&=&
\langle B|A\rangle,\\
\langle A|B\oplus C\rangle
&=&
\langle A|B\rangle\oplus \langle A|C\rangle,\\
\langle A|\Lambda\odot B\rangle
&=&
\Lambda\odot \langle A|B\rangle, \quad \Lambda\in \CC.
\ee

\section{Fourier transform}

Let $A:\mathbb{X}\to \CC$.
The Fourier transform
$\hat A:\mathbb{X}\to \CC$ is defined by
\be
\hat A(K)&=& \Big(\hat A_1(K),\hat A_2(K)\Big)
=
\hat A_1(K)\oplus i'\hat A_2(K)
\\
&=&
\int_{\ominus T\oslash 2'}^{T\oslash 2'} A(X)\odot \Exp(\ominus i' K\odot X)DX.
\ee
After some computations one finds its equivalent explicit form
\be
\hat A(K)
&=&
f^{-1}\left(\Re\int_{-f(T)/2}^{f(T)/2}a(x)e^{-if(K)x}dx\right)
\nonumber\\
&\pp=&
\oplus i'
f^{-1}\left(\Im\int_{-f(T)/2}^{f(T)/2}a(x)e^{-if(K)x}dx\right).
\ee
Now assume $f(T)<\infty$. Dirac's delta in the space of square-integrable functions $a:\mathbb{C}\to \mathbb{C}$, $\int_{-f(T)/2}^{f(T)/2}|a(x)|^2dx<\infty$, can be written as
\be
\delta(x-y)
&=&
\frac{1}{f(T)}\sum_{n\in \mathbb{Z}} e^{i 2n\pi (x-y)/f(T)}\\
&=&
\frac{1}{f(T)}+
\frac{2}{f(T)}\sum_{n>0}
\Bigg(
\cos \frac{2n\pi x}{f(T)}\cos \frac{2n\pi y}{f(T)}
\nonumber\\
&\pp=&
%\pp{\frac{1}{f(T)}+\frac{2}{f(T)}\sum_{n>0}}
+
\sin \frac{2n\pi x}{f(T)}\sin \frac{2n\pi y}{f(T)}
\Bigg).
\ee
Denoting
\be
c_n(y)
&=&
\sqrt{\frac{2}{f(T)}}\cos \frac{2n\pi y}{f(T)},\quad n>0\\
s_n(y)
&=&
\sqrt{\frac{2}{f(T)}}\sin \frac{2n\pi y}{f(T)},\quad n>0\\
c_0(y)
&=&
\sqrt{\frac{1}{f(T)}},\\
s_0(y)
&=&
0,\\
C_n(X) &=& f^{-1}\Big(c_n\big(f(X)\big)\Big),\\
S_n(X) &=& f^{-1}\Big(s_n\big(f(X)\big)\Big),\ee
one finds
\be
\delta(x-y)
&=&
\sum_{n\geq 0}\Big(c_n(x)c_n(y)+s_n(x)s_n(y)\Big),
\ee
which implies
\be
A(X)
&=&
\oplus_{n\geq 0}
\Big(
C_n(X)\odot \langle C_n|A\rangle
\oplus
S_n(X)\odot \langle S_n|A\rangle
\Big).
\nonumber\\
\ee
Since
\be
\langle C_n|C_m\rangle
&=&
f^{-1}\left(\Re\langle c_n|c_m\rangle\right)
\oplus
i'f^{-1}\left(\Im\langle c_n|c_m\rangle\right)\\
&=&
f^{-1}\big(\delta_{nm}\big)
\oplus
i'f^{-1}\left(0\right)\\
&=&
\delta'_{nm}\oplus i' 0'\\
&=&
\delta'_{nm}\\
&=&
\langle S_n|S_m\rangle,\\
\langle C_n|S_m\rangle
&=&
0',
\ee
where
\be
\delta'_{nm}
&=&
f^{-1}(\delta_{nm})=\left\{
\begin{array}{cll}
1' & \textrm{for} & n=m\\
0' & \textrm{for} & n\neq m
\end{array}
\right.
,
\ee
we arrive at the Parseval formula
\be
\langle A|B\rangle
&=&
\int_{\ominus T\oslash 2'}^{T\oslash 2'} A(X)^*\odot B(X)DX
\\
&=&
\oplus_{n\geq 0}
\Big(
\langle A|C_n\rangle\odot \langle C_n|B\rangle
\oplus
\langle A|S_n\rangle\odot \langle S_n|B\rangle
\Big).\nonumber\\
\ee
\section{Example: Cantorian sawtooth function}

Let us consider the sawtooth function $a:\mathbb{R}\to \mathbb{R}$ and its ternary Cantor-line analogue, $A=f^{-1}\circ a\circ f$, depicted in Fig.~2.
\begin{figure}\label{Fig2}
\includegraphics[width=8 cm]{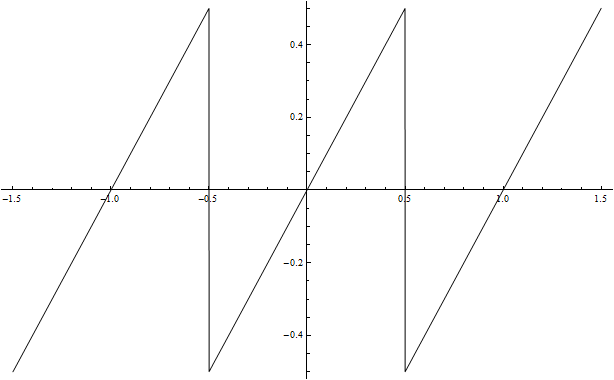}
\includegraphics[width=8 cm]{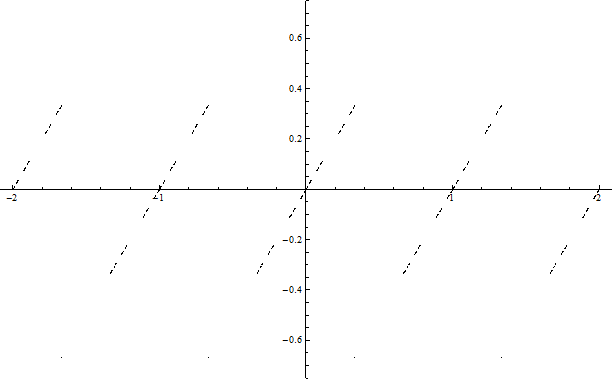}
\caption{The sawtooth function $a$ (upper), and its Cantorian analogue $A=f^{-1}\circ a\circ f$, where $f$ is the ternary Cantor-line function (lower).}
\end{figure}
Now let us perform the Fourier transform with $f(T)=1$, i.e. $T=1'$. Fig.~3 shows two reconstructions of $A$ with 5 and 30 Fourier terms, respectively. The Gibbs phenomenon is clearly visible.
\begin{figure}\label{Fig2}
\includegraphics[width=8 cm]{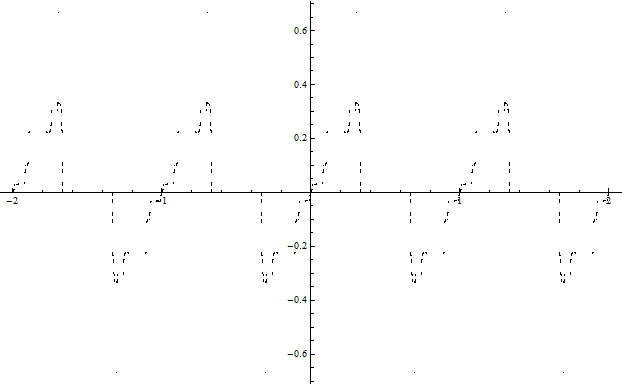}
\includegraphics[width=8 cm]{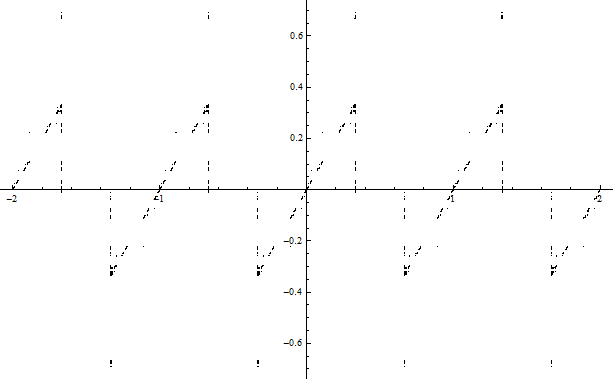}
\caption{Finite-sum Fourier reconstructions of the sawtooth function, with 5 (upper) and 30 terms. The Gibbs effect is enhanced by jumps of the Cantor-line function.}
\end{figure}
\section{Spectrum of frequencies}

The Laplacian $\Delta = \frac{D}{DX}\frac{D}{DX}$ satisfies
\be
\Delta C_n(X)
&=&
\frac{D}{DX}\frac{D}{DX}
f^{-1}\Big(c_n\big(f(X)\big)\Big)\\
&=&
f^{-1}\left(-\frac{(2n\pi)^2}{f(T)^2}\sqrt{\frac{2}{f(T)}}\cos \frac{2n\pi f(X)}{f(T)}\right)\nonumber\\
&=&
\ominus
f^{-1}(n)^{2'}\odot f^{-1}\left(\frac{(2\pi)^2}{f(T)^2}\right)\odot
C_n(X).
\ee
Spectrum in the sense of Jorgensen and Pedersen \cite{JP} corresponds to $f(T)=1$ and is given by $\lambda$s that parametrize the exponent $e^{2\pi i\lambda x}$. For the quaternary Cantor set they find that $\lambda$s are all the odd multiples of $4^j$, for nonnegative integers $j$. In our case the relevant numbers are $\{n'=f^{-1}(n)\}$.
For the quaternary Cantor set $\mathbb{X}_+$ the bijection $f^{-1}$ maps $n=b_k 2^k+\dots+b_1 2^1+b_0$ into $n'=2(b_k 4^k+\dots+b_1 4^1+b_0)$. Now assume that $j$ is the smallest value of the binary index satisfying $b_j=1$. Then
\be
n' &=& 2(b_k 4^k+\dots+b_{j+1} 4^{j+1}+4^j)\\
&=& 4^j2(b_k 4^{k-1}+\dots+b_{j+1} 4^1+1).
\ee
So our $n'$ are given by even multiples of $4^j$, but not all of them are allowed. For example, $4^j\times 6$ is not included.

As opposed to the approach from \cite{JP} one can obtain an analogous spectrum for the ternary Cantor set, and actually for all Cantor sets described in \cite{ACK}, including those that are not self-similar. The ternary Cantor-line function (Fig. 1, upper) satisfies $f^{-1}(k)=k$ for integer $k$, so spectrum will include all non-negative integers. If one takes a self-similar ``middle-third" Cantor set, then $n'=2(b_k 3^k+\dots+b_1 3^1+b_0)$, so one will obtain an analogous result as for the quaternary Cantor set, but with $4^j$ replaced by $3^j$.

\section{Summary}

The concepts of differentiation and integration can be easily defined for fractals equipped with intrinsic non-Diophantine arithmetic. Once we know how to integrate and differentiate in a way that preserves the fundamental theorems of calculus, we can easily define Fourier transforms that possess all the standard properties (resolution of unity, Parceval theorem, Gibbs effect,...). Accordingly, there is no problem with momentum representation in quantum mechanics on fractal space-times, at least in the class of fractals that fulfill our assumptions.

\widetext

\section*{Appendix: Properties of $\langle A|B\rangle$}

Let
\be
\langle A|B\rangle
&=&
f^{-1}\left(\Re\int_{-f(T)/2}^{f(T)/2}
\overline{a(x)}b(x)
dx\right)
\oplus
i'f^{-1}\left(\Im\int_{-f(T)/2}^{f(T)/2}
\overline{a(x)}b(x)
dx\right)
\ee
Then
\be
\langle A|B\rangle^*
&=&
f^{-1}\left(\Re\int_{-f(T)/2}^{f(T)/2}
\overline{a(x)}b(x)
dx\right)
\ominus
i'f^{-1}\left(\Im\int_{-f(T)/2}^{f(T)/2}
\overline{a(x)}b(x)
dx\right)\\
&=&
f^{-1}\left(\Re\int_{-f(T)/2}^{f(T)/2}
\overline{a(x)}b(x)
dx\right)
\oplus
i'f^{-1}\left(-\Im\int_{-f(T)/2}^{f(T)/2}
\overline{a(x)}b(x)
dx\right)\\
&=&
f^{-1}\left(\Re\int_{-f(T)/2}^{f(T)/2}
\overline{b(x)}a(x)
dx\right)
\oplus
i'f^{-1}\left(\Im\int_{-f(T)/2}^{f(T)/2}
\overline{b(x)}a(x)
dx\right)\\
&=&
\langle B|A\rangle
\ee
Now let $A=f^{-1}\circ a\circ f$, $B=f^{-1}\circ b\circ f$. We first show that $A\oplus B=f^{-1}\circ(a+b)\circ f$:
\be
A\oplus B(X)
&=&
A(X)\oplus B(X)\\
&=&
f^{-1}\Big(f\big(A(X)\big)+f\big(B(X)\big)\Big)
\\
&=&
f^{-1}\big(a[f(X)]+b[f(X)]\big)
\\
&=&
f^{-1}\big((a+b)[f(X)]\big)
\\
&=&
f^{-1}\circ(a+b)\circ f(X).
\ee
So
\be
\langle A|B\oplus C\rangle
&=&
f^{-1}\left(\Re\int_{-f(T)/2}^{f(T)/2}
\overline{a(x)}(b+c)(x)
dx\right)
\oplus
i'f^{-1}\left(\Im\int_{-f(T)/2}^{f(T)/2}
\overline{a(x)}(b+c)(x)
dx\right)
\ee
Let us concentrate on the first integral
\be
f^{-1}\left(\Re\int_{-f(T)/2}^{f(T)/2}
\overline{a(x)}(b+c)(x)
dx\right) \label{1int}
\ee
since the proof for
\be
f^{-1}\left(\Im\int_{-f(T)/2}^{f(T)/2}\overline{a(x)}(b+c)(x)dx\right)\label{2int}
\ee
will be identical.
We find
\be
(\ref{1int})
&=&
f^{-1}\left(\Re\int_{-f(T)/2}^{f(T)/2}
\overline{a(x)}b(x)
dx
+
\Re\int_{-f(T)/2}^{f(T)/2}
\overline{a(x)}c(x)
dx\right)
\\
&=&
f^{-1}\left[f\Bigg(f^{-1}\Big(\Re\int_{-f(T)/2}^{f(T)/2}
\overline{a(x)}b(x)
dx\Big)\Bigg)
+
f\Bigg(f^{-1}\Big(\Re\int_{-f(T)/2}^{f(T)/2}
\overline{a(x)}c(x)
dx\Big)\Bigg)\right]
\\
&=&
f^{-1}\left(\Re\int_{-f(T)/2}^{f(T)/2}
\overline{a(x)}b(x)
dx\right)
\oplus
f^{-1}\left(\Re\int_{-f(T)/2}^{f(T)/2}
\overline{a(x)}c(x)
dx\right).
\ee
Analogously
\be
(\ref{2int})
&=&
f^{-1}\left(\Im\int_{-f(T)/2}^{f(T)/2}
\overline{a(x)}b(x)
dx\right)
\oplus
f^{-1}\left(\Im\int_{-f(T)/2}^{f(T)/2}
\overline{a(x)}c(x)
dx\right),
\ee
implying
\be
\langle A|B\oplus C\rangle &=& \langle A|B\rangle \oplus \langle A|C\rangle.
\ee
Next, let $\Lambda\in \CC$ be a constant. Then
\be
B'(X)
&=&
\big(B_1'(X),B_2'(X)\big)\\
&=&
\Lambda\odot B(X)\\
&=&
(\Lambda_1\oplus i'\Lambda_2)\odot (B_1(X)\oplus i'B_2(X))\\
&=&
\Big(\Lambda_1\odot B_1(X)\ominus\Lambda_2\odot B_2(X))\Big)
\oplus
i'\Big(\Lambda_1\odot B_2(X)\oplus\Lambda_2\odot B_1(X))\Big)\\
&=&
f^{-1}\Big(f(\Lambda_1)b_1[f(X)]-f(\Lambda_2)b_2[f(X))]\Big)
\oplus
i'f^{-1}\Big(f(\Lambda_1)b_2[f(X)]+f(\Lambda_2)b_1[f(X)]\Big)
\ee
\be
B'
&=&
\big(B_1',B_2'\big)\\
&=&
\Big(f^{-1}\circ \Big(f(\Lambda_1)b_1-f(\Lambda_2)b_2\Big)\circ f
,
f^{-1}\circ \Big(f(\Lambda_1)b_2+f(\Lambda_2)b_1\Big)\circ f\Big)\\
&=&
\Big(f^{-1}\circ b'_1\circ f
,
f^{-1}\circ b'_2\circ f\Big)\\
&=&
\Big(f^{-1}\circ \Re\Big(\big(f(\Lambda_1)+i f(\Lambda_2)\big)\big(b_1+i b_2\big)\Big)\circ f
,
f^{-1}\circ \Im\Big(\big(f(\Lambda_1)+i f(\Lambda_2)\big)\big(b_1+i b_2\big)\Big)\circ f
\Big)
\ee
So
\be
b'_1
&=&
\Re\Big(\big(f(\Lambda_1)+i f(\Lambda_2)\big)\big(b_1+i b_2\big)\Big),\\
b'_2
&=&
\Im\Big(\big(f(\Lambda_1)+i f(\Lambda_2)\big)\big(b_1+i b_2\big)\Big),\\
b'
&=& b'_1+ib'_2\\
&=&
\big(f(\Lambda_1)+i f(\Lambda_2)\big)\big(b_1+i b_2\big)
\ee
Recall that
\be
\langle A|B'\rangle
&=&
f^{-1}\left(\Re\int_{-f(T)/2}^{f(T)/2}
\overline{a(x)}b'(x)
dx\right)
\oplus
i'f^{-1}\left(\Im\int_{-f(T)/2}^{f(T)/2}
\overline{a(x)}b'(x)
dx\right)
\ee
Therefore
\be
\langle A|\Lambda\odot B\rangle
&=&
f^{-1}\left(\Re\int_{-f(T)/2}^{f(T)/2}
\overline{a(x)}b'(x)
dx\right)
\oplus
i'f^{-1}\left(\Im\int_{-f(T)/2}^{f(T)/2}
\overline{a(x)}b'(x)
dx\right)\\
&=&
f^{-1}\left(\Re\left[\big(f(\Lambda_1)+i f(\Lambda_2)\big)\int_{-f(T)/2}^{f(T)/2}
\overline{a(x)}b(x)
dx\right]\right)
\nonumber\\
&\pp=&\oplus
i'f^{-1}\left(\Im \left[\big(f(\Lambda_1)+i f(\Lambda_2)\big)\int_{-f(T)/2}^{f(T)/2}
\overline{a(x)}b(x)
dx\right]\right)\\
&=&
f^{-1}
\left(
f(\Lambda_1)\Re\int_{-f(T)/2}^{f(T)/2}\overline{a(x)}b(x)dx
-
f(\Lambda_2)\Im\int_{-f(T)/2}^{f(T)/2}\overline{a(x)}b(x)dx
\right)
\nonumber\\
&\pp=&\oplus
i'f^{-1}
\left(
f(\Lambda_2)\Re\int_{-f(T)/2}^{f(T)/2}\overline{a(x)}b(x)dx
+
f(\Lambda_1)\Im\int_{-f(T)/2}^{f(T)/2}\overline{a(x)}b(x)dx
\right)
\\
&=&
f^{-1}
\left(
f(\Lambda_1)f\left[f^{-1}\left(\Re\int_{-f(T)/2}^{f(T)/2}\overline{a(x)}b(x)dx\right)\right]
-
f(\Lambda_2)f\left[f^{-1}\left(\Im\int_{-f(T)/2}^{f(T)/2}\overline{a(x)}b(x)dx\right)\right]
\right)
\nonumber\\
&\pp=&\oplus
i'f^{-1}
\left(
f(\Lambda_2)f\left[f^{-1}\left(\Re\int_{-f(T)/2}^{f(T)/2}\overline{a(x)}b(x)dx\right)\right]
+
f(\Lambda_1)f\left[f^{-1}\left(\Im\int_{-f(T)/2}^{f(T)/2}\overline{a(x)}b(x)dx\right)\right]
\right)
\nonumber
\\
&=&
\Lambda_1\odot f^{-1}\left(\Re\int_{-f(T)/2}^{f(T)/2}\overline{a(x)}b(x)dx\right)
\ominus
\Lambda_2\odot f^{-1}\left(\Im\int_{-f(T)/2}^{f(T)/2}\overline{a(x)}b(x)dx\right)
\nonumber\\
&\pp=&\oplus
i'
\left[
\Lambda_2\odot f^{-1}\left(\Re\int_{-f(T)/2}^{f(T)/2}\overline{a(x)}b(x)dx\right)
\oplus
\Lambda_1\odot f^{-1}\left(\Im\int_{-f(T)/2}^{f(T)/2}\overline{a(x)}b(x)dx\right)
\right]
\\
&=&
(\Lambda_1\oplus i'\Lambda_2)
\odot
\left[
f^{-1}\left(\Re\int_{-f(T)/2}^{f(T)/2}\overline{a(x)}b(x)dx\right)
\oplus i'
f^{-1}\left(\Im\int_{-f(T)/2}^{f(T)/2}\overline{a(x)}b(x)dx\right)
\right]\\
&=&
\Lambda
\odot
\langle A|B\rangle
\ee
which ends the proof.

\end{document}